\documentclass[11pt]{article}

\usepackage{amsmath}
\usepackage[applemac]{inputenc}

\begin{document}

\begin{flushright} ULB-TH/07-18
\end{flushright}
\vspace{.7cm}
\begin{centering}
\textbf{\large{A Note on Conserved Charges of Asymptotically Flat and Anti-de Sitter Spaces in Arbitrary Dimensions}}\\
\vspace{.7cm}

Ella JAMSIN\footnote{FRIA-FNRS bursar (National Fund for Scientific Research, Belgium)}\\
\vspace{.3cm}
\emph{Physique Th\'eorique et Math\'ematique,}\\
\emph{Universit\'e Libre de Bruxelles \& International Solvay Institutes,}\\
\emph{Campus Plaine C.P. 231}\\
\emph{B-1050 Bruxelles, Belgium}\\
\texttt{ejamsin@ulb.ac.be}

\end{centering}
\vspace{.7cm}

\begin{abstract}
The calculation of conserved charges of black holes is a rich problem, for which many methods are known. Until recently, there was some controversy on the proper definition of conserved charges in asymptotically anti-de Sitter (AdS) spaces in arbitrary dimensions. This paper provides a systematic and explicit Hamiltonian derivation of the energy and the angular momenta of both asymptotically flat and asymptotically AdS spacetimes in any dimension $D\geq4$. This requires as a first step a precise determination of the asymptotic conditions of the metric and of its conjugate momentum. These conditions happen to be achieved in ellipsoidal coordinates adapted to the rotating solutions. The asymptotic symmetry algebra is found to be isomorphic either to the Poincaré algebra or to the $\textbf{so}(D-1,2)$ algebra, as expected. In the asymptotically flat case, the boundary conditions involve a generalization of the parity conditions, introduced by Regge and Teitelboim, which are necessary to make the angular momenta finite. The charges are explicitly computed for Kerr and Kerr-AdS black holes for arbitrary $D$ and they are shown to be in agreement with thermodynamical arguments.
\end{abstract}

\newpage

\tableofcontents

\section{Introduction}

Energy is a subtle issue in general relativity. Indeed, in accordance with the equivalence principle, the gravitational contribution to energy cannot be localized. Nevertheless, in the presence of asymptotic conditions, the total energy, as well as angular momentum, of the system can be defined. However, even in such a case, some confusion used to exist in the literature : there are many different methods and in some cases they did not all give the same charges. For example, as has already been emphasized \cite{loi1}, not even in four dimensions did all authors obtain the same expression for the energy of Kerr-AdS black holes and some of these expressions were in disagreement with the first law of black hole thermodynamics. 

Nowadays, though, number of methods give reliable results. In the paper \cite{loi1}, Gibbons et al. compute the energy of Kerr-AdS black holes first indirectly by integrating the first law and then using the conformal definition of Ashtekar, Magnon and Das \cite{AMD1,AMD2}, and they show that both of them agree. Besides the Regge-Teitelboim method \cite{RetT} adopted here, some of the various other definitions are the following : the approach of Abbott and Deser \cite{AD1,AD2,AD3}, the spinor definition \cite{sp1,sp2} based on the electric Weyl tensor, covariant phase space methods \cite{cov1,cov2,cov3,cov4}, cohomological techniques \cite{co1,co2}, the KBL approach \cite{KBL1,KBL2,KBL3}, Noether methods \cite{N1,N2,N3,N4,N5,N6}, the ``counterterm subtraction method" \cite{c1,c2} (and more references on this in \cite{cov3} for example), improved surface integrals \cite{GG} and regularisation of the Euclidean action \cite{Ol1,Ol2}.

In view of the contradictory results that existed in the literature, our main purpose in this paper is to define the conserved quantities through a method that relates unambiguously charges to symmetries without having to make arbitrary choices\footnote{Except for overall additive constants that can be fixed either by background adjustment or using the algebra} and hence stands on a firm footing from this point of view. This method is the Hamiltonian approach introduced by Regge and Teitelboim \cite{RetT} for asymptotically flat spacetimes in dimension $D=4$ and Cartesian coordinates. It associates a conserved charge to any asymptotic, i.e. not necessarly exact, Killing vector. The charges are expressed as surface integrals over a $2$-sphere at infinity and reproduce, as it should, the ADM energy and angular momentum \cite{ADM}.

This paper's aim is the generalisation of the method to arbitrary dimensions for both asymptotically flat and AdS spacetimes and the explicit calculation of the corresponding rotating solution. This requires as a first step a precise definition of asymptotically flat or AdS spacetimes as asymptotic conditions on the metric and its conjugate momenta. Let us focus for a while on asymptotically flat spacetimes although the following generalizes to the asymptotically AdS case. In $D$ spacetime dimensions, one expects the metric describing a rotating mass to differ from the flat metric by terms that decay generically as $r^{3-D}$ as this is the asymptotic behavior of the elementary solution of the Poisson equation. In standard polar coordinates
$$
ds^2=-dt^2+dr^2+r^2 d\Omega^2_{D-2}+h_{\mu\nu}dx^\mu dx^\nu
$$
this would become (denoting any angle by $\hat\alpha, \hat\beta$)
\begin{eqnarray}
h_{tt},h_{tr}, h_{rr} &\sim&r^{3-D} ,\nonumber\\
h_{t\hat{\alpha}},h_{r\hat{\alpha}}&\sim& r^{4-D},\nonumber\\
h_{\hat{\alpha}\hat{\beta}} &\sim& r^{5-D}.\nonumber
\end{eqnarray}
If one compares, however, the Kerr metric in the ellipsoidal Boyer-Lindquist coordinates to the flat metric in standard spherical coordinates, one finds, for example, terms of order $\frac{1}{r^2}$ in $h_{rr}$, which are unacceptable when $D>5$. If inserted in the standard expressions giving the conserved charges, these terms would lead to divergences. Fortunately, this difficulty can be cured. One must either transform the black hole metric into new coordinates such that the expected decays are fulfilled,  or one must rewrite the formalism in coordinates adapted to the $\frac{1}{r^2}$ terms, i.e. in coordinates in which the $\frac{1}{r^2}$ terms can be included in the background. The two approaches are equivalent because the techniques are covariant (as long as one expresses the background and the deviation in the same system of coordinates). For convenience, we consider here the second method and develop the formalism in order to admit ellipsoidal systems of coordinates.

Notice that the vanishing mass limit of the Kerr metric written in ellipsoidal coordinates depends on the parameters describing the rotation of the black hole. The question then arises whether it makes sense to compare the charges of two different black holes (i.e. with different values of those parameters) while the two background metrics seem to be different. The answer turns out to be yes because, for any value of the parameters, the background is the Minkowski metric and all charges vanish when the mass vanishes.

The generalisation of the method to asymptotically AdS spacetimes in dimension $D=4$ and the explicit calculation of the energy and angular momentum of Kerr-AdS black holes was made by Henneaux and Teitelboim in \cite{HetT}. The generalisation of the method to higher dimensions was then given in \cite{MH} in standard spherical coordinates but the computation of charges of Kerr-AdS black holes in arbitrary dimension was not included as the corresponding metric had not been derived yet. This is now the case \cite{deSit}, and we show in this paper that the Hamiltonian method gives charges in agreement with the thermodynamical arguments of \cite{loi1}.

In \textbf{section \ref{ham}}, we recall the main points and formulas of the Hamiltonian formulation of gravitation. Then we derive the surface integrals defining the conserved charges by requiring the functional derivatives of the Hamiltonian with respect to canonical variables to be well defined.

\textbf{Section \ref{flat}} is devoted to asymptotically flat spacetimes. Their definition imposes asymptotic conditions on the components of the metric and their conjugate momenta. In particular, it is shown that besides the fall-off conditions the parity conditions introduced in four dimensions must be generalized in higher dimensions. Using these conditions, we give a simpler expression for the charges. We then briefly discuss the asymptotic symmetries of asymptotically flat spacetimes and the Poisson bracket algebra of the charges. We then focus on the Kerr metric in arbitrary dimension and compute its charges explicitly.

The same work is presented for asymptotically AdS spacetimes in \textbf{section \ref{ads}}. The asymptotic conditions are then simpler as we do not need to introduce any parity condition.

As Hamiltonian methods are not universally used, in order to be self contained, this paper contains a wide part of review material.

\section{Hamiltonian Formulation and Charges}
\label{ham}
In this section, we briefly review the essential formulas of the Hamiltonian formulation of gravitation and derive the conserved charges as integrals over a $D-2$-sphere at infinity. A complete description of Hamiltonian formulation of gravitation can be found in \cite{ADM,Dirac,CHS,Wald,MTW}. In what follows, Greek indices range from $0$ to $D$ while latin indices range from $1$ to $D$, the comma denotes the usual derivative and $/$ is the covariant derivative with respect to the metric $g_{\mu\nu}$.

\subsection{Hamiltonian formulation}

The Hamiltonian formulation of a field theory requires a breakup of spacetime into space and time. The lapse function $N$ and the $D-1$ components of the shift vector $N^i$ are related to the metric components by 
\begin{equation}
\begin{split}
\label{NNi}
N &= (-g^{00})^{-\frac{1}{2}}\\
N_i &= g_{0i}.
\end{split}
\end{equation}

We then work with the $\{ N, N^i ,g_{ij}\}$ instead of the $\{ g_{\mu\nu}\}$, only the $g_{ij}$ being canonical fields. Their conjugate momenta $\pi^{ij} = \frac{\delta L}{\delta \dot g_{ij}}$ are given by
\begin{equation}
\label{piij}
\pi^{ij}=-\sqrt g(K^{ij}-Kg^{ij})
\end{equation}
where $K_{ij}=(2N)^{-1}(-g_{ij}+N_{i/j}+N_{j/i})$ is the extrinsic curvature, $K = K^a_{\ a}$ its trace and $g=\det(g_{ij})$.

It can be shown that the Hamiltonian is then given by
\begin{equation}
H[g_{ij}, \pi^{ij}] = \int \mathrm d^{D-1}x\{N(\textbf x)\mathcal{H}(\textbf x)+N^i(\textbf x)\mathcal{H}_i(\textbf x)\} + \mathrm{boundary\ term}
\end{equation}
where 
\begin{equation}
\begin{split}
\label{HHi}
&\mathcal{H} = g^{-1/2} (\pi_{ij}\pi^{ij}-\frac{1}{D-2}\pi^2) - g^{1/2}R+2\Lambda g^{1/2}\\
&\mathcal{H}_i = -2\pi_{i \ /j} ^{\ j}.
\end{split}
\end{equation}
The boundary term (that is a surface integral over a $D-2$-dimensional closed surface at infinity) is fixed by requiring that the functional derivatives of the Hamiltonian with respect to the canonical variables are well defined \cite{RetT}. 

The vacuum Einstein equations $G_{\mu\nu} = 0$ are equivalent to the system formed by the Hamiltonian equation and the constraints $\mathcal{H}=0 , \mathcal{H}_i=0$.

\subsection{Conserved charges}

More generally, the deformation defined by the asymptotic Killing vector field $\xi = \xi^\bot\textbf n + \xi^i\textbf e_i$ are generated, in the canonical formalism, by \cite{MH}
\begin{equation}
\label{Q}
Q[\xi] = Q_0[\xi]+ \mathcal{I} [\xi] =  \int \mathrm d^{D-1}x \{\xi^\bot(\textbf x)\mathcal{H}(\textbf x) + \xi^i(\textbf x)\mathcal{H}_i(\textbf x)  \} + \mathcal{I}[\xi]
\end{equation}
where $\mathcal{I}[\xi]$ is the boundary term. 

For a general Killing vector, the generator $Q[\xi]$ does not vanish when the constraints are taken into account. It then reduces to $\mathcal I[\xi]$, that consequently defines the conserved charge associated to the Killing vector field $\xi(\textbf x)$. 

As already mentioned, the surface integral $\mathcal I[\xi]$ is determined by requiring that $Q[\xi] =Q[g_{ij}, \pi^{ij}; \xi]$ has well defined functional derivatives. In other words, its variation must be given by a volume integral only :
\begin{equation}
\label{del}
\delta Q[\xi]=\int \mathrm d^{D-1}x\{A^{ij}(\textbf x) \delta g_{ij}(\textbf x) + B_{ij}(\textbf x) \delta \pi^{ij}(\textbf x)\}.
\end{equation}

Using the explicit form of $\mathcal H$ and $\mathcal H_i$ (\ref{HHi}), one can compute the variation of the volume integral $Q_0[\xi]$ in equation (\ref{Q}). One finds (from \cite{RetT} generalized to $D$ dimensions) 
\begin{eqnarray}
\label{deltaQ}
\delta Q_0[\xi] & = & \int \mathrm d^{D-1}x\ \{A^{ij}(\textbf x) \delta g_{ij}(\textbf x) + B_{ij}(\textbf x)\delta\pi^{ij}(\textbf x)\} \nonumber \\
&& -\oint \mathrm d^{D-2}s_l\ G^{ijkl} ( \xi^\bot\delta g_{ij/k} - \xi^\bot_{\ ,k}\delta g_{ij}) \nonumber \\
&& -\oint \mathrm d^{D-2}s_l\ \{2\xi_k \delta \pi^{kl} + (2\xi^k \pi^{jl} - \xi^l \pi^{jk}) \delta g_{jk}\}.
\end{eqnarray}
where
\begin{equation}
\label{Gijkl}
G^{ijkl} \equiv \frac{1}{2} \sqrt{g}(g^{ik}g^{jl}+g^{il}g^{jk}-2g^{ij}g^{kl})
\end{equation}
$A^{ij}(\textbf x)$ and $B_{ij}(\textbf x)$ are the functional derivatives of $Q[\xi]$. Their explicit form is not necessary here. The surface integrals are taken over a sphere at spatial infinity : $r\rightarrow\infty$.
 
 From (\ref{Q}), (\ref{del}) and (\ref{deltaQ}), we get that 
 \begin{eqnarray}
 \delta \mathcal I[\xi] &=& \oint \mathrm d^{D-2}s_l\ G^{ijkl} ( \xi^\bot\delta g_{ij/k} - \xi^\bot_{\ ,k}\delta g_{ij}) \nonumber\\
  \label{dI}
 &+& \oint \mathrm d^{D-2}s_l\ \{2\xi_k \delta \pi^{kl} + (2\xi^k \pi^{jl} - \xi^l \pi^{jk}) \delta g_{jk}\}.
 \end{eqnarray} 
The next step is then to rewrite the right hand side of (\ref{dI}) as the variation of something. In order to achieve this, we need the asymptotic behavior of the metric. We will restrict ourselves to two specific cases, namely asymptotically flat and asymptotically AdS spacetimes.

\section{Asymptotically Flat Spacetimes}

In this chapter, asymptotically flat spacetimes will be precisely defined by a series of boundary conditions on the metric and its conjugate momenta. The general formula for conserved Poincaré charges will then be given. We will apply it to the case of the Kerr metric in arbitrary dimension which describes a rotating black hole. This chapter generalizes in higher dimensions and more general coordinates the work of Regge and Teitelboim \cite{RetT}.

\label{flat}
\subsection{Definition and conditions}
Asymptotically flat spacetimes are defined as spacetimes that approach the Minkowski one at large distance. We accordingly consider metrics of the form
\begin{equation}
\label{g}
g_{\mu\nu} = \bar g_{\mu\nu} + h_{\mu\nu}
\end{equation}
where $\bar g_{\mu\nu}$ is the Minkowski background metric and the perturbation $h_{\mu\nu}$ tends to zero at spatial infinity : $h_{\mu\nu}\rightarrow 0$ when $r\rightarrow \infty$. The asymptotic symmetry group is Poincar\' e. Moreover, the fall-off law of the perturbation must be specified such that it obeys the following criteria :

(i) it should be invariant under the action of the Poincar\' e group since otherwise a symmetry transformation would map an allowed configuration onto a non-allowed one;

(ii) it should make the surface integrals associated with the generators of Poincar\'e finite;

(iii) it should include asymptotically flat solutions of physical interest, such as the Kerr metric.

The asymptotic behavior of the perturbation can be firstly bounded by using the fact that the linearized approximation of general relativity should be valid at spatial infinity. By imposing the invariance of this behavior under the Poincar\' e transformations, one then finds asymptotic conditions for the momenta. In what follows, we will always consider systems of polar coordinates composed of one time variable $t$, one radial variable $r$ and $D-2$ dimensionless variables, i.e. angles or functions of angles. The latter will be denoted $\hat\alpha$, $\hat\beta$, ... In this notation, the whole set of conditions reads
\begin{eqnarray}
h_{tt} &=& h_{tt}^{(0)}r^{3-D} + \mathcal{O}(r^{2-D}),\nonumber\\
h_{tr} &=& h_{tr}^{(0)}r^{3-D} + \mathcal{O}(r^{2-D}),\nonumber\\
h_{t\hat{\alpha}} &=& h_{t\hat{\alpha}}^{(0)}r^{4-D} + \mathcal{O}(r^{3-D}),\nonumber\\
h_{rr} &=& h_{rr}^{(0)}r^{3-D} + \mathcal{O}(r^{2-D})\nonumber,\\ 
h_{r\hat{\alpha}} &=& h_{r\hat{\alpha}}^{(0)}r^{4-D} + \mathcal{O}(r^{3-D})\nonumber,\\
\label{hij}
h_{\hat{\alpha}\hat{\beta}} &=& h_{\hat{\alpha}\hat{\beta}}^{(0)}r^{5-D} + \mathcal{O}(r^{4-D}),\\
&& \nonumber\\
\label{pirr}
\pi^{rr} &=& \pi^{rr}_{(0)}+\mathcal{O}(r^{-1})\nonumber,\\
\pi^{r\hat{\alpha}} &=& \pi^{r\hat\alpha}_{(0)}r^{-1}+\mathcal{O}(r^{-2})\nonumber,\\
\pi^{\hat{\alpha}\hat{\beta}} &=& \pi^{\hat\alpha\hat\beta}_{(0)}r^{-2}+\mathcal{O}(r^{-3}).
\end{eqnarray}
Notice that in polar coordinates the asymptotic conditions on the momenta are independent of the dimension.

These conditions lead to infinite angular momenta in all dimension : they are given by surface integrals over a sphere at spatial infinity whose integrand's leading term grows like $r$. One way to understand this is to observe that the charges associated with translations are finite. These charges involve the first order term in the deviation and the translation Killing fields, which go to constant at infinity. This implies that the leading term in the angular momenta will diverge since the corresponding Killing vectors grow like $r$ at infinity. This divergence can be avoided if extra conditions are imposed to make the divergent term vanish. However, we cannot require the fields to decay faster at spatial infinity since it would exclude the Kerr solution. The idea of Regge and Teitelboim \cite{RetT}, inspired by the knowledge of the metric of the rotating black holes, is to impose parity conditions under space inversion $\textbf x\rightarrow -\textbf x$. They can be summarized in the following way :
\begin{eqnarray}
h_{ij}^{(0)}dx^idx^j  && \ \ \mathrm{is\ even\ under\ inversion,}\\
\label{par}
\pi^{ij}_{(0)}\frac{\partial}{\partial x^i}\frac{\partial}{\partial x^j}  && \ \ \mathrm{is\ odd\ under\ inversion.}
\end{eqnarray}
Consequently, $h_{ii}^{(0)}$ must be even and $\pi^{ii}_{(0)}$ odd. For the other components, $r \hat\alpha$ or $\hat\alpha \hat\beta$, it depends on the parity of the chosen dimensionless coordinate. 

Let us notice that the asymptotic conditions constrain the coordinates, as will soon become clear.

\subsection{Conserved charges}
If we use the conditions (\ref{g}) to (\ref{par}), we can show that the right hand side of (\ref{dI}) can be rewritten as
\begin{equation}
\delta \big\{\oint d^{D-2}s_r\ (\bar g^{ik}\bar g^{jr}-\bar g^{ij}\bar g^{kr})( \xi^\bot h_{ij;k}- \xi^\bot_{\ ,k} h_{ij} )
+ \ 2\oint d^{D-2}s_r\  \bar\pi^r_{\ k}\xi^k\big\},
\end{equation}
where the semi-colon is the covariant derivative with respect to the flat background metric $\bar g_{ij}$ and $\bar\pi^{ij}$ is the momentum where $g_{ij,0}$ is replaced by $h_{ij,0}$ and everywhere else the total metric $g_{ij}$ is replaced by the flat metric $\bar g_{ij}$. In particular, covariant derivatives must only be calculated with respect to the flat background.

Consequently, $\mathcal I\left[\xi\right]$ is known within an additive constant. We fix it so that all charges vanish for Minkowski spacetime. The conserved charge associated to the Killing vector field $\xi(\textbf x)$ is then given by
\begin{equation}
\label{I}
\mathcal I\left[\xi\right] = \oint d^{D-2}s_r\ (\bar g^{ik}\bar g^{jr}-\bar g^{ij}\bar g^{kr}) (\xi^\bot h_{ij;k}- \xi^\bot_{\ ,k} h_{ij} ) + 2\oint d^{D-2}s_r\  \bar\pi^r_{\ k}\xi^k.
\end{equation}

\subsection{Asymptotic symmetries}

It can be shown (see the appendix for more details) that the most general deformation preserving the asymptotic flatness is not only Poincar\' e but rather 
$$
\xi^\mu(\textbf x) = \xi^\mu_P(\textbf x) + \xi^\mu_+(\textbf x)
$$
where $\xi^\mu_P(\textbf x) = A^\mu_{\ \nu}x^\nu+B^\mu$, with $A^\mu_{\ \nu}$ antisymmetric, is the generator of Poincar\' e transformations and $\xi^\mu_+(\textbf x)$ is a function falling off as $r^{4-D}$ and odd under inversion at leading order. Consequently, the conditions above are invariant under an infinite-dimensional group containing the Poincar\' e group, and denoted here by $G$. However, the additional transformations $\xi^\mu_+(\textbf x)$ form an invariant subgroup $H$ of $G$. Moreover, they do not contribute to the surface integrals (\ref{dI}) so that the the physical symmetry group is the factor group $G/H$, that is, the Poincar\' e group. Notice that in four dimensions, the parity conditions, first introduced to make all conserved charges finite, also play a role in the vanishing of the charges associated to $\xi_+$ and the consistent factorization of $G/H$ to the Poincaré group while in higher dimension the fall-off conditions are enough for the latter issues.

\subsection{Poisson bracket algebra}

We now show that, due to our choice of  the additive constant, the Poisson bracket of the charges is just isomorphic to the Lie algebra of the infinitesimal asymptotic symmetries, i.e. that 
\begin{equation}
\{Q\left[\xi\right],Q\left[\lambda\right]\}=Q\left[\left[\xi,\lambda\right]\right].
\end{equation}
In general, the charges only yield a projective representation of the asymptotic symmetry group \cite{HB1,HB2} :
\begin{equation}
\label{cent}
\{Q\left[\xi\right],Q\left[\lambda\right]\}=Q\left[\left[\xi,\lambda\right]\right] + K\left[\xi,\lambda\right].
\end{equation}
In (\ref{cent}), the central charges $K\left[\xi,\lambda\right]$ do not involve the canonical variables. On might rewrite (\ref{cent}) as 
\begin{equation}
\label{centbis}
\delta_{\lambda}Q\left[\xi\right]=Q\left[\left[\xi,\lambda\right]\right] + K\left[\xi,\lambda\right].
\end{equation}
Let us evaluate (\ref{centbis}) on the flat background. The left hand side vanishes because the asymptotic symmetries are exact symmetries of the background. The first term of the right hand side vanishes by our choice of the additive constant. Consequently, $K\left[\xi,\lambda\right]$ vanishes on the background. As it does not depend on the metric, the central charge is identically zero with our adjustment of the integration constants.

\subsection{Application to Kerr metric}

Myers and Perry derived first in Cartesian coordinates the metric of the most general rotating black hole in any spacetime dimension \cite{MetP}. However, in that system of coordinates, the parity conditions (\ref{par}) are not respected. Consequently, we have to find a system of coordinates in which the asymptotic conditions are fulfilled\footnote{More precisely, the procedure involves a background substraction (see (\ref{g}). In principle, it is covariant if one transforms not only the perturbation but also the background under a coordinate transformation. However, the new background might take an awkward form. If one insists on simple background components, the systems of coordinates are restricted. }. Such a system is given by the Boyer-Lindquist (B-L) coordinates, which are ellipsoidal coordinates. This characterization will become clearer further in the text. The description of the metric is a review of \cite{MetP}.

In order to treat both odd and even dimensional cases at the same time, let us set $D=2n+1+\epsilon$ where $\epsilon = 0$ if $D$ is odd and  $\epsilon = 1$ if $D$ is even. The B-L coordinates are $(t,r,\mu_i, \phi_j)$ where $i = 1, \dots, n+\epsilon$ and $j =\nolinebreak 1,\dots,n$. Here, $t$ is the time variable, $r$ is the radial variable and the $\mu_i, \phi_j$ are $D-1$ dimensionless variables. The latter are not all independant, the $\mu_i$ are related to each other by the relation $\sum_{i=1}^{n+\epsilon}\mu_i^2 = 1$. In these coordinates, the Kerr metric in arbitrary number of dimensions is given by
\begin{eqnarray}
\label{BL}
ds^2&=&-dt^2 + \sum_{i=1}^{n+\epsilon}(r^2+a_i^2)d\mu_i^2+\sum_{i=1}^{n}(r^2+a_i^2)\mu_i^2d\phi_i^2\nonumber\\
&&+\frac{\mu r^{2-\epsilon}}{\Pi F} (dt+\sum_{i=1}^{n} a_i\mu_i^2d\phi_i)^2+\frac{\Pi F}{\Pi - \mu r^{2-\epsilon}}dr^2,
\end{eqnarray} 
where
\begin{equation}
\begin{split}
&F=1-\sum_{i=1}^{n}\frac{a_i^2\mu_i^2}{r^2+a_i^2},\\
&\Pi = \prod_{i=1}^{n} (r^{2}+a_{i}^{2}).
\end{split}
\end{equation}
In our notation, when $D$ is even, $a_{n+1}=0$. The flat background metric is the limit of this metric for $\mu$ tending to $0$\footnote{This is obvious by considering the Kerr metric in cartesian coordinates, given by equations (3.9) and (3.10) or (3.12) of \cite{MetP}} : $d\bar s^2 = \lim_{\mu\rightarrow0}ds^2$. The remaining terms make up the perturbation. Consequently, in the B-L coordinates\footnote{More rigorously, in a system of coordinates that asymptotically coincides with the Boyer-Lindquist ones. Indeed, the way we defined the flat background in B-L coordinates is not exactly the same as applying the change of coordinates on the flat metric from cartesian coordinates, but both definitions coincide asymptotically.}, the flat metric is 
\begin{equation}
d\bar s^2=-d{t}^2 + Fdr^2 + \sum_{i=1}^{n+\epsilon}(r^2+a_i^2)d\mu_i^2+\sum_{i=1}^n(r^2+a_i^2)\mu_i^2d\phi_i^2.
\end{equation}
The adjective ``ellipsoidal'' can be understood by noticing that this is the Minkowski metric $d\bar s^2 = -dt^2 + \sum_{i=1}^n(dx_i^2+dy_i^2)+\epsilon dz^2$ in an unusual system of coordinates, defined from the Cartesian coordinates by 
\begin{eqnarray} 
\label{cartBL}
x_i&=&(r^2 + a_i^2)^{1/2}\mu_i \cos  \phi_i  \nonumber \\
y_i&=&(r^2 + a_i^2)^{1/2}\mu_i \sin  \phi_i \\
z&=&r\mu_{n+\epsilon}.\nonumber
\end{eqnarray}
($i=1, \dots, n$). Unless the $a_i$ parameters are all zero, the $(r, \mu_i, \phi_j)$ are not spherical coordinates\footnote{To be precise, for $D$ odd, the coordinates are spherical as soon as the parameters are all equal, not necessarily zero.}. Moreover, the radial coordinate $r$ is implicitly defined by using (\ref{cartBL}) in the relation between the $\mu_i$'s :
\begin{equation}
\label{rel}
\sum_{i=1}^{n}\frac{x_i^2+y_i^2}{r^{2}+a_{i}^{2}}+\epsilon\frac{z^2}{r^2}=1.
\end{equation}
For arbitrary rotation parameters, this is the equation of an ellipsoid of revolution. 

A natural question arising then is whether we can get rid of the terms of $d\bar s^2$ depending on the $a_i$'s by regarding them as part of the perturbation. This would be a way to write the metric in spherical coordinates and corresponds to defining the flat background metric as the limit of the Kerr metric when all parameters are zero (not only $\mu$). We actually can see that the asymptotic conditions are not fulfilled by those terms for $D>5$. For example, consider the $a_i$ term in $\bar g_{\mu_i\mu_i}$. It is of order $1$. On the other hand, from (\ref{hij}) we see that $h_{\mu_i\mu_i}$ must decrease like $r^{5-D}$. Consequently, as soon as $D>5$, that term cannot be considered as a perturbation\footnote{Using the definition of $F$, one also finds the problematic $\frac{1}{r^2}$ term in $h_{rr}$ that is used as an example in the introduction.}. 

Nevertheless, we can easily define spherical coordinates $(t, \hat r, \hat\mu_i, \hat\phi_j)$ in which the asymptotic conditions are satisfied by imposing 
\begin{equation}
\label{sphell}
\hat r^2\hat\mu_i^2 = (r^2 + a_i^2)\mu_i^2 \ \ \ \  , \ \ \ \  \hat\phi_j = \phi_j.
\end{equation} 
The metric is then a bit more complicated because it has more terms in the perturbation. Nevertheless, it actually does not change the calculation of conserved charges at all because neither the terms that are removed nor the ones that are added contribute to the surface integrals.

The explicit computation of the Poincaré charges of the Kerr black hole in any dimension can then be performed. They are obtained by explicitly writing the Killing vector $\xi$ in (\ref{I}) for each Poincaré transformation. 

By using the asymptotic behavior of the elements in $\mathcal I\left[\xi\right] $, one can show that the charges associated to the translations and the boosts vanish. A few more symmetry arguments imply that the only non vanishing components of the angular momentum are the ones associated to the rotations parameterized by the  $n=\left[\frac{D-1}{2}\right]$ angles $\phi_i$, and they will be noted $L_i$. This result could be predicted using that the Cartan subalgebra of $so(D-1)$ is the direct sum of $n$ $\textbf u(1)$ algebras, each one acting as a rotation of one of the angles $\phi_i$. 

The energy $E$, associated to the time translations, that are generated by $\xi = \frac{\partial}{\partial t}=\textbf n$, also does not vanish.

By inserting the components of the metric in the surface integral, solving a few technical problems and then reintroducing the gravitational constant, we get 
\begin{equation}
\begin{split}
E &=  \frac{(D-2) A_{D-2}}{16\pi G}\mu,\\
L_i &= \frac{A_{D-2}}{8\pi G}\mu a_i,
\end{split}
\end{equation}
where $ A_{D-2}$ is the volume of the unit $D-2$-sphere, given by $$A_{D-2} = \frac{2\pi^{\frac{D-1}{2}}}{\Gamma(\frac{D-1}{2})}.$$
These results are in accordance with the charges calculated by Myers and Perry in \cite{MetP}. They lead to the following interpretation of the parameters of the metric \nolinebreak : $\mu$ is a measure of the mass of the system and the $\mu a_i$ give the components of the angular momentum. Although these results are well-known, it is interesting to see that we recover them by using this Hamiltonian method.

\section{Asymptotically AdS Spacetimes}
\label{ads}

As done for the flat case, this section will first give a precise description of asymptotically AdS spacetimes, leading to a simpler formula for the conserved charges, as presented in \cite{MH}. The formula will then be applied to the rotating AdS black hole (the so-called Kerr-AdS black hole).

\subsection{Definition and conditions}

Anti-de Sitter spacetime is the maximally symmetric vacuum solution of the Einstein equations, with a negative cosmological constant. Its group of motions is $O(D-1,2)$. It is described by the following metric :
\begin{equation}
\label{bads}
d\bar s^2 = -(1-\lambda \hat r^2) dt^2 + (1-\lambda \hat r^2)^{-1}+\hat r^2d\Omega_{D-2}
\end{equation}
where the cosmological constant is $\Lambda = (D-1)\lambda$ and $d\Omega_{D-2}$ is the surface element of the unit $D-2$-sphere. 

Similar to the flat case, asymptotically AdS spacetimes are defined as spacetimes approaching AdS at large distances,  with a set of boundary conditions fulfilling the following requirements :

(i) the conditions should be invariant under the action of $O(D-1,2)$;

(ii) the surface integrals associated with the generators of $O(D-1,2)$ should be finite;

(iii) the asymptotic conditions should include the asymptotically AdS solutions of physical interest, such as the Kerr-AdS metric\footnote{The Kerr-AdS metric will be given in section \ref{kads}}.

It turns out \cite{MH} that these criteria can be fulfilled by demanding that the metric deviations $h_{\mu\nu}$ from the AdS background ($g_{\mu\nu} = \bar g_{\mu\nu} + h_{\mu\nu}$) and the momenta $\pi^{ij}$ conjugate to the spatial metric behave asymptotically as follows (using the same coordinate notation as for the flat case) :

\begin{eqnarray}
h_{tt} &=& h_{tt}^{(0)}r^{3-D} + \mathcal{O}(r^{2-D})\nonumber\\
h_{tr} &=& h_{tr}^{(0)}r^{-D} + \mathcal{O}(r^{-1-D})\nonumber\\
h_{t\hat{\alpha}} &=& h_{t\hat{\alpha}}^{(0)}r^{3-D} + \mathcal{O}(r^{2-D})\nonumber\\
h_{rr} &=& h_{rr}^{(0)}r^{-1-D} + \mathcal{O}(r^{-2-D})\nonumber\\ 
h_{r\hat{\alpha}} &=& h_{r\hat{\alpha}}^{(0)}r^{-D} + \mathcal{O}(r^{-1-D})\nonumber\\
\label{hijads}
h_{\hat{\alpha}\hat{\beta}} &=& h_{\hat{\alpha}\hat{\beta}}^{(0)}r^{3-D} + \mathcal{O}(r^{2-D})\\
&& \nonumber\\
\label{pirrads}
\pi^{rr} &=& \pi^{rr}_{(0)}r^{-1}+\mathcal{O}(r^{-2})\nonumber\\
\pi^{r\hat{\alpha}} &=& \pi^{r\hat\alpha}_{(0)}r^{-2}+\mathcal{O}(r^{-3})\nonumber\\
\pi^{\hat{\alpha}\hat{\beta}} &=& \pi^{\hat\alpha\hat\beta}_{(0)}r^{-5}+\mathcal{O}(r^{-6}).
\end{eqnarray}

Some comments are in order. First, by replacing the above behavior in (\ref{dI}), notice that in the AdS case, we do not need to add any parity conditions to get finite surface integrals. Let's also emphasize the slight difference from Henneaux's paper \cite{MH} in the way we consider the conditions on the momenta. There, they are introduced as a consequence of the conditions on the metric deviation and the specific form (\ref{bads}) of the background AdS metric. In this paper, on the other hand, we would like to be able to consider also systems of coordinates in which the AdS background is not given by (\ref{bads}) but includes more terms. We should then consider both the behavior of the deviation and the behavior of the momentum as conditions to be verified. 

\subsection{Conserved charges}

Using the behavior (\ref{pirrads}), it can be shown that in this case the surface integral giving the conserved charges is 
\begin{equation}
\label{Iads}
\mathcal I[\xi] = \oint d^{D-2}s_r\ (\bar g^{ik}\bar g^{jr}-\bar g^{ij}\bar g^{kr}) (\xi^\bot h_{ij;k}- \xi^\bot_{\ ,k} h_{ij} ) + 2\oint d^{D-2}s_r\  \bar\pi^r_{\ k}\xi^k.
\end{equation}
where the semi-colon denotes covariant differentiation in the spatial AdS background $\bar g_{ij}$. The additive constant is fixed so that the charges vanish on the AdS background.

\subsection{Asymptotic symmetries}

The $O(D-1,2)$ transformations are not the only ones that conserve the boundary conditions above. It can be seen \cite{MH} that the whole set of asymptotic symmetries form an infinite Lie algebra. Nevertheless, as they do not change the surface integral $\mathcal I[\xi]$, the extra transformations are pure gauge transformations and can then be consistently factored out. Once this is done, only the finite-dimensional AdS algebra is left. From the physical point of view, this algebra is accordingly the asymptotic symmetry algebra\footnote{In the case when $D=3$, the situation is quite different. You can find more details about it in \cite{MH}.}.

\subsection{Poisson bracket algebra}

The argument given in the asymptotically flat case to show the vanishing of the central charges directly generalizes to asymptotically AdS spacetimes for $D>4$\footnote{For $D=3$ the central charge does not vanish as shown in \cite{HB2}.}. 

\subsection{Application to Kerr-AdS metric}
\label{kads}
Gibbons, Lü, Page and Pope constructed in \cite{deSit} the general Kerr-(anti-)de Sitter metric in arbitrary spacetime dimension $D\geq4$, that is, the most general solution for a rotating black hole in asymptotically (anti-)de Sitter spacetime. We restrict ourselves to the negative cosmological constant  case because, contrary to the positive one, its spacelike surfaces are open (non-compact) and it has an asymptotic structure.

The ellipsoidal system of coordinates $(t, r, \mu_i, \phi_i)$ used in this case generalizes the one used in the flat case to a non zero cosmological constant but we will still call it Boyer-Lindquist. It is related to the spherical coordinates $(t, \hat r, \hat\mu_i, \hat\phi_i)$ by 
\begin{equation}
\label{sphellb}
\hat r^2\hat\mu_i^2 = \frac{r^2 + a_i^2}{1+\lambda a_i^2}\mu_i^2 \ \ \ \  , \ \ \ \  \hat\phi_j = \phi_j.
\end{equation}
Compare to (\ref{sphell}). In such a system of coordinates, the AdS metric (\ref{bads}) becomes (still using the $\epsilon$ notation introduced in the flat case)
\begin{eqnarray}
d\bar s^2&=& -W(1-\lambda r^2)dt^2 + \frac{U}{V}dr^2+\sum_{i=1}^n\frac{r^2+a_i^2}{1+\lambda a_i^2}\mu_i^2d\phi_i^2\nonumber\\
&&+\sum_{i=1}^{n+\epsilon}\frac{r^2+a_i^2}{1+\lambda a_i^2}d\mu_i^2+\frac{\lambda}{W(1-\lambda r^2)}(\sum_{i=1}^{n+\epsilon}\frac{r^2+a_i^2}{1+\lambda a_i^2}\mu_id\mu_i)^2,
\end{eqnarray}
where
\begin{eqnarray}
W&=&\sum_{i=1}^{n+\epsilon}\frac{\mu_i^2}{1+\lambda a_i^2},\nonumber\\
U&=&r^\epsilon\sum_{i=1}^{n+\epsilon}\frac{\mu_i^2}{r^2+a_i^2}\prod_{b=1}^{n}(r^2+a_b^2),\nonumber\\
V&=&r^{\epsilon-2}(1-\lambda r^2)\prod_{b=1}^{n}(r^2+a_b^2).\nonumber
\end{eqnarray}
The Kerr-AdS metric, depending on $n+1$ parameters $M$, $a_i$, is given by
\begin{equation}
ds^2=d\bar s^2 + \frac{2M}{U}(Wdt-\sum_{i=1}^n\frac{a_i\mu_i^2}{1+\lambda a_i^2}d\phi_i)^2 + \frac{2MU}{V(V-2M)}dr^2.
\end{equation}
It is straightforward to see that, in these coordinates, the metric deviations $h_{\mu\nu}$ have the right fall-off. It takes little writing to check that the momenta are good too. 

The change of coordinates (\ref{sphellb}) can be used to go back to spherical coordinates. The background is then simpler but the deviation includes more terms. In any event, the asymptotic conditions are still fulfilled. 

The only non vanishing charges are again found to be the energy $E$ associated to $\xi_E = \frac{\partial}{\partial t}$ and the $n$ components $L_i$ of the angular momentum associated to the rotations generated by $\xi_{\bar R_i}=\frac{\partial}{\partial \phi_i}$. In this case, no charge can be shown to vanish only by looking at the fall-off of the associated Killing vector and the gravitational variables. Nevertheless, by using the AdS Killing vectors eplicitly calculated in \cite{Kil} and adapting them to our $(t, r, \phi_i, \theta_j)$ coordinates, it is easily seen that all undesired components depend on a function of the integration variables whose symmetry is such that the surface integrals vanish. 

It is useful, first, to notice that all $\bar\pi^r_{\ k}$ momenta vanish except $\bar\pi^r\phi_i$. Consequently, we can focus only on components $\xi^\bot$ and $\xi^{\phi_i}$ of the Killing vectors. 

For $D$ even, for all AdS Killing vectors but $\xi_E$ and $\xi_{\bar R_i}$ these components depend on $\sin\phi_k$ or $\cos\phi_k$ for some $k$ : hence, the associated charges are proportionnal either to $\int_0^{2\pi}\sin\phi_kd\phi_k$ or to $\int_0^{2\pi}\cos\phi_kd\phi_k$ that both vanish. 

For $D$ odd, the same thing happens for all vectors except a few of them. Nevertheless, the latter are prorportional to $\cos\theta_{n+1}$. As all gravitational variables are invariant for $\theta_{n+1}\rightarrow\pi-\theta_{n+1}$ and as the integration is made for $\theta$ from $0$ to $\pi$, it vanishes. 

The energy is associated to the Killing vector $\xi_E = \frac{\partial}{\partial t}$. But in this case, $\xi^2\neq-1$ so that $\xi^\bot=N\neq1$ ($N$ is the lapse function, defined in equation (\ref{NNi})). We accordingly have to keep one more term in the energy than for the asymptotically flat case : 
\begin{equation}
E= \oint d^{D-2}s_l\ (\bar g^{ik}\bar g^{jl}-\bar g^{ij}\bar g^{kl}) (Nh_{ij;k}-N_{,k} h_{ij} ).
\end{equation}
The angular momenta have the same form as in the asymptotically flat case \nolinebreak:
\begin{equation}
L_i = 2\oint d^{D-2}s_l\  \bar\pi^l_{\ k}\xi^k.
\end{equation}

I started my calculation of the conserved charges of the Kerr-AdS black hole in the spherical coordinates, going back to ellipsoidal coordinates later to simplify the integrals. The latter were finally computed using Mathematica for $d=4, \dots, 11$. 

When the dust settles, after reintroducing the gravitational constant, one finds
\begin{eqnarray}
E&=&\frac{MA_{D-2}}{4\pi G \prod_{j=1}^n(1+\lambda a_j^2)}(\sum_{i=1}^n\frac{1}{1+\lambda a_i^2}-\frac{1-\epsilon}{2}),\\
L_i&=&\frac{MA_{D-2}}{4\pi G \prod_{j=1}^n(1+\lambda a_j^2)}\frac{a_i}{(1+\lambda a_i^2)}.
\end{eqnarray}
These results can be checked to be in accordance with (4.12), (4.15) and (4.16) of \cite{loi1}. Consequently, they satisfy the first law of thermodynamics.

\section{Conclusion}

In this paper, we generalized the Regge-Teitelboim approach of charges in general relativity to asymptotically flat and AdS spacetimes in arbitrary dimension $D>4$ and ellipsoidal coordinates. In particular we provided generalized explicit boundary conditions, including parity conditions in the asymptotically flat case. It was shown that the charges satisfy the algebra of the asymptotic symmetries. We then applied the method to rotating black holes in both cases and found expressions that are in agreement with the thermodynamics of black holes. In view of the recently existing controversy about the proper definitions of charges in the case under consideration, we have tried to be as explicit as possible.

Recently, AdS with a scalar field has been much studied. It makes necessary to relax the asymptotic conditions (\ref{hijads}) and (\ref{pirrads}) to more general ones \cite{cov3,C1,Wis,C2,C3,HM1,HM2,C4,AM}. It would be interesting to study how our method can be adapted to that case and to investigate the thermodynamics of such spacetimes.

\section*{Acknowledgements}
I would like to thank Marc Henneaux for advice and guidance throughout this work. I am also grateful to Glenn Barnich and Geoffrey Comp\` ere for useful discussions and to Chethan Krishnan for proofreading. This work is supported in part by a ``Pôle d'Attraction Interuniversitaire" (Belgium), by IISN-Belgium, convention 4.4505.86, by the National Fund for Scientific Research (FNRS Belgium) and by the European Commission programme MRTN-CT-2004-005104, in which I am associated to V.U. Brussels.

\appendix

\section{Derivation of the asymptotic symmetries}
\subsection{Description of the problem}

In Cartesian coordinates, the $D$-dimensional phase space of general relativity is defined, in the case of asymptotically flat spacetimes, by the following conditions :
\begin{equation}
\label{condash}
g_{ab}(\textbf{x}) = \eta_{ab} + h_{ab}(\textbf x),
\end{equation}
\begin{equation}
\begin{split}
h_{ab}(\textbf x)=&h_{ab}^{(0)}(\textbf x)r^{3-D}+\mathcal O(r^{2-D})\\
h_{ab}^{(0)}(\textbf x)=&h_{ab}(-\textbf x)
\end{split}
\end{equation}
et
\begin{equation}
\begin{split}
\label{condaspi}
\pi^{ab}(\textbf x)=&\pi^{ab}_{(0)}(\textbf x)r^{2-D} + \mathcal O(r^{1-D})\\
\pi^{ab}_{(0)}(\textbf x)=&-\pi^{ab}(-\textbf x).
\end{split}
\end{equation}

Under a deformation of the spacelike hypersurface on which the canonical variables $g_{ab}(\textbf{x})$ and $\pi^{ab}(\textbf x)$ are defined, the variation of these variables is given by \cite{MTW}
\begin{eqnarray}
\label{deltag}
\delta g_{ab}&=&2\xi^0G_{abcd}\pi^{cd} + \xi_{a/b} + \xi_{b/a}\\
\label{deltapi}
\delta \pi^{ab}&=&-\xi^0g^{1/2}(R^{ab}-\frac{1}{2}g^{ab}R) + \frac{1}{2}\xi^0g^{-1/2}g^{ab}(tr\pi^2-\frac{1}{N-1}(tr\pi)^2)\nonumber\\
&&-2\xi^0g^{-1/2}(\pi^{ai}\pi^{bj}-\frac{1}{N-1}\pi^{ab}g^{ij})\pi_{ij}\nonumber\\
&&+g^{1/2}(\xi^{0/ab}-g^{ab}\xi^{0/i}_{\ \ \ /i})+(\pi^{ab}\xi^{i})_{/i}\nonumber\\
&&-\xi^a_{\ /i}\pi^{ib}-\xi^b_{\ /i}\pi^{ia}
\end{eqnarray}
where  $(\xi^0(\textbf x), \xi^k(\textbf x))$ are the parameters caracterising the deformation, $G_{abcd}$ is defined at equation (\ref{Gijkl}), $/$ is the covariant derivative with respect to $g_{ab}$ and $tr$ denotes the matrix trace. In this appendix, the most general parameters $(\xi^0(\textbf x), \xi^k(\textbf x))$ preserving the asymptotic conditions (\ref{condash}) to (\ref{condaspi}) are found. This is done in two steps : \\

\begin{enumerate}

\item  One first looks for the most general deformation preserving the asymptotic conditions in the particular case of the flat configuration : $g_{ab}(\textbf x) = \delta_{ab}$, $\pi^{ab}(\textbf x) = 0$. One thus look for the most general $(\xi^0(\textbf x), \xi^k(\textbf x))$ such that
\begin{eqnarray}
\label{deltagplat}
\delta (g_{ab}=\delta_{ab})&=&\xi_{a,b} + \xi_{b,a}=\mathcal O(r^{3-D}),\ \mathrm {even,}\\
\label{deltapiplat}
\delta (\pi^{ab}=0) &\propto&\xi^{0,ab}-\delta^{ab}\Delta\xi^0 =\mathcal O(r^{2-D}), \ \mathrm{odd,}
\end{eqnarray}
where $\Delta \xi^0 = \delta^{ab} \xi^0 _{\ ,ab}$ is the Laplacian with respect to the flat metric of $\xi^0$ and the parity conditions are demanded on the leading order of each expression. \\

\item One can then check that the solution of (\ref{deltagplat}) and (\ref{deltapiplat}) is actually also the most general solution in the case of the most general configuration verifying (\ref{condash}) to (\ref{condaspi}).\\

Let us assume that $\xi^0$ and $\xi^a$ can be developed in positive and negative powers of $r$, as for example,
\begin{equation}
\label{dev}
\xi^0 = a_nr^n + a_{n-1}r^{n-1} +\dots+a_1r+a_0+a_{-1}+\dots.
\end{equation}
where the coefficients $a_j$ depend on the angles $\beta$ on the $(D-2)$-sphere. In particular, let us assume that $\xi^0$ et $\xi^a$ do not depend on $\log r$. One then writes $\xi^0=\mathcal O(r^n)$ if $a_k=0$ for $k>n$.\\
\end{enumerate}
\subsection{Transformation of the flat configuration}

\emph{Fall-off conditions :}\\
\\
Let us first consider the fall-off conditions. One has to solve
\begin{eqnarray}
\label{defplatea}
\xi_{a,b} + \xi_{b,a}&=&\mathcal O(r^{3-D}),\\
\label{defplateb}
\xi^0_{\ ,ab}-\delta_{ab}\Delta\xi^0 &=&\mathcal O(r^{2-D}).
\end{eqnarray}
To study these equations, one will need the following lemma :\\
\textbf{Lemma : } \emph{Let $f$ be a function that can be developed as (\ref{dev}) and such that
\begin{equation}
\partial_if = \mathcal O(r^{-s})\ \ \ ,\ \ \ s\geq1
\end{equation}
where $\partial_i = \frac{\partial}{\partial x^i}$. Then
\begin{eqnarray}
&f=\mathcal O(r^{-s+1})+C&\ \ \ \ \ \ \ s>1\\
&f=\mathcal O(1) & \ \ \ \ \ \ \ s=1
\end{eqnarray}
where $C$ is a constant.}

\textbf{Proof : } On first writes the conditions in terms of polar coordinates $\beta = \{\beta_1, \beta_2, \dots\}$ 
\begin{eqnarray}
\partial_rf &=& \frac{g_r(\beta)}{r^s}+\mathcal O(r^{-s-1})\\
\partial_{\beta_j} f &=& \frac{g_{\beta_j}(\beta)}{r^{s-1}}+\mathcal O(r^{-s}).
\end{eqnarray}
As $f$ exists, the integrability conditions are satisfied. On $g_r$ and $g_{\beta_j}$ they read $\partial_{\beta_j}g_r =$ \newline$-(s-1)g_{\beta_j}$, $\partial_{\beta_j}g_{\beta_k}=\partial_{\beta_k}g_{\beta_j}$. 
\begin{description}
\item[$s>1$ :] By integrating over $r$, on has
\begin{equation}
f = -\frac{g_r(\beta)}{(s-1)r^{s-1}}+C(\beta) + \mathcal O(r^{-s}).
\end{equation}
One must add the condition $\partial_{\beta_j}C=0$, otherwise $\partial_{\beta_j}f$ would not behave as $r^{-s+1}$. One then gets $f=\mathcal O(r^{-s+1})+C$ as expected.
\item[$s=1$ :] In this case, $\partial_{\beta_j}g_r = 0$ for any angle $\beta_j$. Accordingly, $g_r = K$ where $K$ is a constant. When integrating over $r$, one finds 
\begin{equation}
f = K\log r +C(\beta_i) + \mathcal O(r^{-1}).
\end{equation}
For $f$ to admit the developpement (\ref{dev}), one must ask $K=0$. One then finds $f=\mathcal O(1)$. 
\end{description}

\textbf{QED}
\\

Using the lemma, one can easily solve (\ref{defplatea}) and (\ref{defplateb}).
\\

Equation (\ref{defplatea}) implies
\begin{equation}
\label{xiabc}
\xi^a_{\ ,bc} = \mathcal O(r^{2-D}).
\end{equation}
Indeed, by deriving each side of (\ref{defplatea}) with respect to $x^c$, one finds
\begin{equation}
\label{xiabcbac}
\xi_{a,bc} + \xi_{b,ac} = \mathcal O(r^{2-D}).
\end{equation}
Besides, by using successively that the partial derivatives with respect to independent variables commute and equation (\ref{defplatea}), one deduces
\begin{eqnarray}
\xi_{b,ac} &=&  \xi_{b,ca}\nonumber\\
&=& -\xi_{c,ba} + \mathcal O(r^{2-D})\nonumber\\
&=&  -\xi_{c,ab} + \mathcal O(r^{2-D})\nonumber\\  
&=&  \xi_{a,cb} + \mathcal O(r^{2-D})\nonumber\\ 
&=&  \xi_{a,bc} + \mathcal O(r^{2-D}).\nonumber 
\end{eqnarray}
When replacing in (\ref{xiabcbac}), one finds (\ref{xiabc}).
\\

Moreover, equation (\ref{defplateb}) implies
\begin{equation}
\label{xi0ab}
\xi^0_{\ ,ab} = \mathcal O(r^{2-D}).
\end{equation}
Indeed, by taking the trace of each side of (\ref{defplateb}), one has
\begin{equation}
(-D+2)\Delta\xi^0=\mathcal O(r^{2-D})
\end{equation}
and as $D\neq2$, it implies that $\Delta \xi^0 = \mathcal O(r^{2-D})$, so that (\ref{xi0ab}).
\\

Let us assume $D\geq 4$, so $D-2 > 1$. From (\ref{xiabc}) and (\ref{xi0ab}), one then finds that 
\begin{eqnarray}
\xi^0_{\ ,a}&=&A^0_{\ a}+\mathcal O(r^{3-D})\\
\xi^a_{\ ,b}&=&A^a_{\ b}+\mathcal O(r^{3-D}).
\end{eqnarray}

To proceed to the next integration, one considers two cases and uses the lemma :

(a) $D\geq5$ : Then, $D-3>1$ and so 
\begin{eqnarray}
\xi^0&=&A_ax^a+B^0 + \xi_+^0(\textbf x)\\
\xi^a&=&A^a_{\ b}x^b+B^a+ \xi_+^a(\textbf x).
\end{eqnarray}
where $\xi_+^0(\textbf x),\xi_+^a(\textbf x)=\mathcal O (r^{4-D})$.

(b) $D=4$ : Then, $D-3=1$ and so
\begin{eqnarray}
\xi^0&=&A_ax^a+ \xi_+^0(\textbf x)\\
\xi^a&=&A^a_{\ b}x^b+ \xi_+^a(\textbf x).
\end{eqnarray}
where $\xi_+^0(\textbf x),\xi_+^a(\textbf x)=\mathcal O (1)$.\\

In order for these solutions of (\ref{xiabc}) et (\ref{xi0ab}) to be also solutions of (\ref{defplatea}) and (\ref{defplateb}), one must demand $\xi_{a,b}+\xi_{b,a}=\mathcal O(r^{3-D})$ and so $A_{ab}=-A_{ba}$.\\
\\
\emph{Parity conditions :}\\
\\
One now has to demand parity conditions for $\delta(\delta_{ab})$ and $\delta(0)$. This implies

(a) $D > 4$ : 
\begin{eqnarray}
\xi_{+(0)}^0(\textbf x) &=& -\xi_{+(0)}^0(-\textbf x)\\
\xi_{+(0)}^a(\textbf x) &=& -\xi_{+(0)}^a(-\textbf x).
\end{eqnarray}

(b) $D=4$ :
\begin{eqnarray}
\xi_{+(0)}^0(\textbf x) &=&B^0 +\bar\xi_{+(0)}^0(\textbf x)\\
\xi_{+(0)}^a(\textbf x) &=&B^a +\bar\xi_{+(0)}^a(\textbf x).
\end{eqnarray}
with
\begin{eqnarray}
\bar\xi_{+(0)}^0(\textbf x) &=& -\bar\xi_{+(0)}^0(-\textbf x)\\
\bar\xi_{+(0)}^a(\textbf x) &=& -\bar\xi_{+(0)}^a(-\textbf x).
\end{eqnarray}

Abandonning the bars, one can summarize the results in the following way : for $D\geq4$, the most general deformation $\xi^\mu = (\xi^0, \xi^i)$ that conserves the asymptotic conditions is given by
\begin{equation}
\xi^\mu=A^\mu_{\ b}x^b+B^\mu + \xi_+^\mu(\textbf x).
\end{equation}
where
\begin{eqnarray}
A_{ab}&=&-A_{ba}\\
\xi_+^\mu(\textbf x) &=& \mathcal O (r^{4-D})\\
\xi_+^\mu(\textbf x) &=& -\xi_+^\mu(-\textbf x).
\end{eqnarray}

Notice that this analysis is done on a given spacelike hypersurface, of equation $x^0=\mathrm{constant}$. By redefining $B^a$, one can write $A^a_{\ b}x^b+B^a=A^a_{\ b}x^b+A^a_{\ 0}x^0+B'^{a}$ with $A_{a0}=-A_{0a}$. The first two terms of the deformation define a Poincaré deformation. To this Poincaré transformation is added another transformation, parameterized by $\xi_+^\mu(\textbf x)$ and happening to be ``pure gauge" as the corresponding surface integrals vanish for any configuration. Notice that for $D=4$, these gauge behave as $\mathcal O(1)$, just like translations. They are sometimes called ``translations depending on the angles". What distinguishes them from real translations is their parity. \\

\subsection{Transformation of the general configuration} 

One still has to check that the solution of (\ref{defplatea}) and (\ref{defplateb}) we have just constructed also conserves the asymptotic conditions (\ref{condash}) to (\ref{condaspi}) in the general case. One accordingly has to check that $\delta g_{ab}$ defined by (\ref{deltag}) behaves as $\mathcal O(r^{3-D})$ and is even at this order and that $\delta \pi^{ab}$ defined by (\ref{deltapi}) behaves as $\mathcal O(r^{2-D})$ and is odd at this order. It can be done term by term rather easily.


\begin{thebibliography}{}
\bibitem{loi1} G. W. Gibbons, M. J. Perry and C. N. Pope, The First Law of Thermodynamics for Kerr-Anti-de Sitter Black Holes, Class. Quant. Grav., 22, 1503 (2005) hep-th/0408217.
\bibitem{AMD1} A. Ashtekar and A. Magnon, Asymptotically anti-de Sitter space-times, Class. Quant. Grav., 1, L39 (1984).
\bibitem{AMD2} A. Ashtekar and S. Das, Asymptotically anti-de Sitter space-times : Conserved quantities, Class. Quant. Grav., 17, L17 (2000) hep-th/9911230.
\bibitem{RetT} T. Regge and C. Teitelboim, Role of Surface Integrals in the Hamiltonian Formulation of General Relativity, Ann. Phys., 88, 286 (1974).
\bibitem{AD1} L. F. Abbott   and S. Deser, Stability of gravity with a cosmological constant, Nucl. Phys., 88, 286 (1974).
\bibitem{AD2} S. Deser, I. Kanik and B. Tekin, Conserved Charges of Higher D Kerr-AdS Spacetimes, Class. Quant. Grav., 22, 3383 (2005).
\bibitem{AD3} S. Deser and B. Tekin, New energy definition for higher-curvature gravities, Phys. Rev., D75, 084032 (2007) gr-qc/0701140.
\bibitem{sp1} G. W. Gibbons, S. W. Hawking, G.T. Horowitz and M. J. Perry, Positive Mass Theorems For Black Holes, Commun. Math. Phys., 88, 295 (1983).
\bibitem{sp2} G. W. Gibbons, C. M. Hull and N. P. Warner, The Stability of Gauged Supergravity, Nucl. Phys. B218, 173 (1983).
\bibitem{cov1} V. Iyer and R. M. Wald, Some properties of Noether charge and a proposal for dynamical black hole entropy, Phys. Rev., D50, 846 (1994) gr-qc/9403028.
\bibitem{cov2} R. M. Wald and A. Zoupas, A general definition of conserved charge and a proposal for dynamical black hole entropy, Phys. Rev., D61, 084027 (2000) gr-qc/9911095.
\bibitem{cov3} S. Hollands, A. Ishibashi and D. Marolf, Comparison between various notions of conserved charges in asymptotically AdS-spacetimes, hep-th/0503045.
\bibitem{cov4}  R. Cai and L. Cao, Conserved charges in even dimensional asymptotically locally anti-de Sitter space-times, JHEP, 0603, 083 (2006) hep-th/0601101.
\bibitem{co1} I. M. Anderson and C. G. Torre, Asymptotic conservation laws in field theory, Phys. Rev. Lett., 77, 4109 (1996) hep-th/9608008.
\bibitem{co2} G. Barnich and F. Brandt, Covariant theory of asymptotic symmetries, conservation laws and central charges, Nucl. Phys. B633, 3 (2002) hep-th/0111246.
\bibitem{KBL1} J. Katz, J. Bicak and D. Lynden-Bell, Relativistic conservation laws and integral constrains for large cosmological perturbations, Phys. Rev., D55, 5957 (1997) gr-qc/0504041.
\bibitem{KBL2} N. Deruelle and J. Katz, On the mass of a Kerr-anti-de Sitter spacetime in d dimensions, Class. Quant. Grav., 22, 421 (2005) gr-qc/0410135.
\bibitem{KBL3} A. N. Aliev, Electromagnetic Properties of Kerr-Anti-de Sitter Black Holes, Phys. Rev. D75, 084041 (2007) hep-th/0702129
\bibitem{N1} B. Julia and S. Silva, Currents and superpotentials in classical gauge invariant theories. I : Local results with applications to perfect fluids and general relativity, Class. Quant. Grav., 15, 2173 (1998) gr-qc/9804029.
\bibitem{N2} S. Silva, On superpotentials and charge algebras of gauge theories, Nucl. Phys., B558, 391-415 (1999) hep-th/9809109.
\bibitem{N3} B. Julia and S. Silva, Currents and superpotentials in classical gauge invariant theories. II : Global aspects and the example of affine gravity, Class. Quant. Grav., 17, 4733 (2000) gr-qc/0005127.
\bibitem{N4} R. Aros, M. Contreras, R. Olea, R. Troncoso and J. Zanelli, Conserved charges for gravity with locally AdS asymptotics, Phys. Rev. Lett., 84, 1647 (2000) gr-qc/9909015.
\bibitem{N5} R. Aros, M. Contreras, R. Olea, R. Troncoso and J. Zanelli, Conserved Charges for Even Dimensional Asymptotically AdS Gravity Theories, Phys. Rev.,  D62, 044002 (2000) hep-th/9912045.
\bibitem{N6} Y. Obukhov and G. F. Rubilar, Invariant conserved currents in gravity theories with local Lorentz and diffeomorphism symmetry, Phys. Rev., D74, 064002 (2006) gr-qc/0608064.
\bibitem{c1} M. Henningson and K. Skenderis, The holographic Weyl anomaly, JHEP, 9807, 023  (1998)hep-th/9806087.
\bibitem{c2} V. Balasubramanian and P. Kraus, A stress tensor for anti-de Sitter gravity, Commun. Math. Phys., 208, 413 (1999) hep-th/9902121.
\bibitem{GG} G. Barnich and G. Comp\` ere, Generalized Smarr relation for Kerr AdS black holes from improved surface integrals, Phys. Rev., D71, 044016 (2005) gr-qc/0412029.
\bibitem{Ol1} R. Olea, Mass, Angular Momentum and Thermodynamics in Four-Dimensional Kerr-AdS Black Holes, JHEP, 0510, 067 (2005) hep-th/059179.
\bibitem{Ol2} R. Olea, Regularisation of odd-dimensional AdS gravity : Kounterterms, JHEP, 0704 073 (2007).
\bibitem{ADM} R. Arnowitt, S. Deser and C. W. Misner, The Dynamics of General Relativity, in Gravitation : an introduction to current research, L. Witten ed., New York, 1962.
\bibitem{HetT} M. Henneaux and C. Teitelboim, Asymptotically Anti-de Sitter Spaces, Commun. Math. Phys., 98, 391 (1985)
\bibitem{MH} M. Henneaux, Asymptotically Anti-de Sitter Universes in $d=3, 4$ and Higher Dimensions, Proceedings of the Fourth Marcel Grossman meetings on General Relativity, R. Ruffini, Elsevier Science Publisher, 1986. 
\bibitem{deSit} G. W. Gibbons, H. Lü, D. N. Page and C. N. Pope, The General Kerr-de Sitter Metrics in All Dimensions, J. Geom. Phys., 53, 49  (2005)hep-th/0404008.
\bibitem{Dirac} P. A. M. Dirac, Fixation of coordinates in the Hamiltonian theory of gravitation, Phys. Rev., 114, 924 (1959).
\bibitem{CHS} A. Hanson, T, Regge and C. Teitelboim, Constrained Hamiltonian systems, 1976, Lecture notes of a lesson cycle in Academia Nazionale dei Lincei, Rome.
\bibitem{Wald} R. M. Wald, General Relativity, University of Chicago Press, Chicago, 1984 
\bibitem{MTW} C. W. Misner, K. S. Thorne and J. A. Wheeler, Gravitation, Freeman, New York, 22nd edition, 1999.
\bibitem{HB1} J. D. Brown and M. Henneaux, On The Poisson Brackets of Differentiable Generators in Classical Field Theory, J. Math. Phys., 27, 489 (1986).
\bibitem{HB2} J. D. Brown and M. Henneaux, Central Charges in the Canonical Realization of Asymptotic Symmetries : An Example from Three Dimensional Gravity, Commun. Math. Phys., 104, 207 (1986).
\bibitem{MetP} R. C. Myers and M. J. Perry, Black holes in higher dimensional Space-Times, Ann. Phys., 172, 304 (1986).
\bibitem{Kil} G. Barnich, F. Brandt and K. Claes, Asymptotically anti-de sitter space-times : symmetries and conservation laws revisited, Nucl. Phys., 127, 114 (2004).
\bibitem{C1} M. Henneaux, C. Martinez, R. Troncoso and J. Zanelli, Black Holes and Asymptotics of 2+1 gravity coupled to a scalar field, Phys. Rev., D65, 104007 (2002) hep-th/0201170.
\bibitem{Wis} E. Winstanley, On the existence of conformally coupled scalar field hair for black holes in (anti-)de Sitter space, Found. Phys., 33, 111 (2003) gr-qc/0205092.
\bibitem{C2} D. Sudarsky and J. A. Gonzalez, On black hole scalar hair in asymptotically anti-de Sitter spacetimes, Phys. Rev., D67, 024038 (2003) gr-qc/0207069.
\bibitem{C3} M. Henneaux, C. Martinez, R. Troncoso and J. Zanelli, Asymptotically Anti-de Sitter spacetimes and scalar fields with a logarithmic branch, Phys. Rev., D70, 044034 (2004) hep-th/0404236.
\bibitem{HM1} T. Hertog and K. Maeda, Black Hole with Scalar Hair and Asymptotics in N=8 Supergravity, JHEP, 0407, 051 (2004) hep-th/0404261. 
\bibitem{HM2} T. Hertog and K. Maeda, Stability and Thermodynamics of AdS Black Holes with Scalar Hair, Phys. Rev., D71, 024001 (2005) hep-th/0409314.
\bibitem{C4} M. Henneaux, C. Mart\'inez, R. Troncoso and J. Zanelli, Asymptotic behavior and Hamiltonian analysis of anti-de Sitter gravity coupled to scalar fields, Annals Phys., 322, 824 (2007) hep-th/0603185.
\bibitem{AM} A. J. Amsel and D. Marolf, Energy bounds in Designer Gravity, Phys. Rev., D74, 064006 (2006) hep-th/0605101.

\end{thebibliography}
\end{document}